\documentclass[11pt,a4paper]{article}

\usepackage[T1]{fontenc}
\usepackage[utf8]{inputenc} 
\usepackage{graphicx}
\usepackage{amsmath,amssymb}
\usepackage{cite}
\usepackage[colorlinks=true,linkcolor=blue,citecolor=blue,urlcolor=blue]{hyperref}
\usepackage{orcidlink}

\usepackage{QCDEX2026}

\begin{document}

\noindent\includegraphics[width=0.4\textwidth]{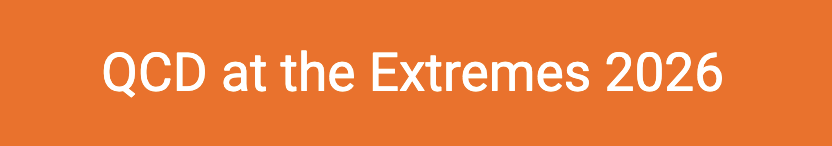}\par
\vspace{0.6em}
\begin{center}
  {\LARGE\bfseries  Determination of the strong coupling constant from the R ratio obtained in KEDR and BESIII experiments\par}
  \vspace{0.8em}
  {\large
    Korneliy Todyshev$^{1,2,*}$\orcidlink{0000-0002-3356-4385}
    \quad and\quad Andrei Kataev$^{3},^{4}$\orcidlink{0000-0002-4006-2498}\par}
  \vspace{0.5em}
  {\small
    $^{1}$ Budker Institute of Nuclear Physics of Siberian Branch Russian Academy of Sciences 11, Acad. Lavrentieva Pr., Novosibirsk, 630090 Russia\par
    $^2$ Novosibirsk State University, 630090, Novosibirsk, Russia\par
    $^{3}$   Institute for Nuclear Research of the Russian Academy of Sciences,
              117312, Moskow, Russia\par
    $^{4}$   Bogoliubov Laboratory of Theoretical Physics, Joint Institute for
     Nuclear Research, 141980, Dubna, Russia \par
    }
  \vspace{0.35em}
  {\small $^{*}$Corresponding author: \href{mailto:todyshev@inp.nsk.su}{todyshev@inp.nsk.su}\par}
  \vspace{0.35em}
  {\small\itshape refereed by the conveners of QCDEX2026\par}
\end{center}
\vspace{0.5em}

\begin{abstract}
The experimental data collected by the KEDR and BESIII collaborations at energies below the charm quark threshold are compared with the massless QCD expressions for the $e^+e^-$ annihilation R-ratio truncated at different orders of perturbation theory. The fits demonstrate the dependence of the 
extracted $\alpha_s(M_Z)$ values on the orders of truncation of the corresponding approximations.
\end{abstract}

\section{Introduction}
\label{sec:intro}
The strong coupling constant is a fundamental parameter determining the
strength of the strong interaction in QCD. It is determined with high
precision from data on high-energy processes,  analyses of quarkonium states, and hadronic $\tau$-lepton decays. 
The relative precision of this parameter, averaged over available
data, is approximately $0.8\%$  at the Z-boson scale. The results of this work are presented in the article \cite{Kataev:2026gea}; 
the primary focus is on determining the strong coupling constant based on the $R$-ratio, as measured in the low-energy region by the KEDR \cite{Anashin:2015woa,Anashin:2016hmv,KEDR:2018hhr} and BESIII \cite{BESIII:2021wib} experiments.
The obtained value of the strong coupling constant can
be compared with results based on data regarding the inclusive
$e^+e^-$ annihilation cross-section into hadrons at low energies, in
order to estimate how well perturbative QCD describes physical processes
in this energy range. 
Research in this energy range has a long history, 
but the experiments BESIII and KEDR, which significantly improved measurement accuracy, 
were performed relatively recently. 
This has heightened interest in this field of energies \cite{Shen:2023,Boito:2025qfz}.
\section{Strong coupling constant $\alpha_s$ and R}
\label{sec:alphasR}
The quantity R, determined experimentally, is directly related to the
strong interaction constant. Within the framework of perturbative QCD, R can be expressed as a series in which the coefficients of the powers of $\alpha_s$ have been calculated up to the fourth order for massless quarks 
and up to the third order for the massive case, and in general form it can be written as
  \begin{eqnarray}
                &&      R^{th}(s)=3\sum Q_f^2\bigg[F_0\bigg(\frac{m_f^2}{s}\bigg)+ F_1\bigg(\frac{m_f^2}{s}\bigg)\frac{\alpha_s(s)}{\pi}+F_2\bigg(\frac{m_f^2}{s}\bigg)\bigg(\frac{\alpha_s(s)}{\pi}\bigg)^2  \nonumber\\
                &+&F_3\bigg(\frac{m_f^2}{s}\bigg)\bigg(\frac{\alpha_s(s)}{\pi}\bigg)^3+F_4(0)\bigg(\frac{\alpha_s(s)}{\pi}\bigg)^4+F_5^{est}(0)\bigg(\frac{\alpha_s(s)}{\pi}\bigg)^5\dots
                \bigg] \nonumber  \\ \nonumber
                &&      +\bigg(\sum_f Q_f\bigg)^2\bigg[\Phi_3\bigg(\frac{m_f^2}{s}\bigg)\bigg(\frac{\alpha_s(s)}{\pi}\bigg)^3+\Phi_4(0)\bigg(\frac{\alpha_s(s)}{\pi}\bigg)^4+\Phi_5^{est}(0)\bigg(\frac{\alpha_s(s)}{\pi}\bigg)^5\dots\bigg]
   \end{eqnarray}
Here functions $F_i$ and $\Phi_i$ correspond to non-singlet and singlet contributions, and the indices assigned to them correspond to powers of $\alpha_s$.
Although the massless approximation formulas were used in our study  \cite{Kataev:2026gea}, the effect of mass on the calculation results was also estimated. 

\section{Fit procedure}
The cited works employed the following minimization function.
\begin{equation}
       \chi^2_{0}= \sum_{ij} \bigg(R^{{exp}}(s_i)-R^{{th}}(s_i)\bigg)
       C^{-1}_{ij} \bigg(R^{{exp}}(s_j)-R^{{th}}(s_j)\bigg) \nonumber
\end{equation}
{\small where $R^{{exp}}(s_i)$ and  $R^{{th}}(s_i)$ are the measured and
theoretically defined values of the  R  ratio, at the concrete energy points,  and
$C^{-1}_{ij}$ are the  coefficients of the inverse covariance matrix.
}
In our analysis we use alternative $\chi^2$ function.
\begin{equation}
\chi^2_1=\frac{(\nu-1)^2}{\nu^2\sigma_0^2} + \frac{1}{\nu^2} \sum_{ij} \bigg(\nu R^{{exp}}(s_i)-R^{{th}}(s_i)\bigg) C_{ij}^{-1} \bigg(\nu R^{{exp}}(s_j)-R^{{th}}(s_j)\bigg) \nonumber
\end{equation}
where $\nu$ is a free normalization parameter, $\sigma_0$ is the minimal correlated systematic uncertainty.\\
A similar function was used, in particular, in \cite{Marshall:1988ri}, but without accounting for correlations between energy points.
Using  the $\chi^2_{1}$ function  allows us to separate the overall bias in
the data and the shape (energy dependence) of the curve that is
actually compared to the theoretical dependence in the fit. 
The reasons for such a shift are primarily associated with possible
systematic inaccuracies in the measurement of luminosity and the
detection efficiency of hadronic events. Formally, one can also use
another minimization function by introducing a scaling factor for the
theoretical calculation of  the R value; the result will be
identical. Both options provide a certain degree of fit flexibility that the
standard function lacks.
\section{Results}
The results of the data fit for KEDR and BESIII is
presented in Table \ref{Tab:Results}. Each column lists the results for the
corresponding order of calculation. 
\renewcommand{\arraystretch}{1.1}
\begin{center}
\begin{table}[h!]
  \caption {\label{Tab:Results} Results of combined fit of KEDR and truncated BESIII data}                                 
\begin{tabular}[c]{|c|c|c|c|c|c|} \hline
R(s) approximation        & $\mathcal{O}(\alpha_s^2)$ & $\mathcal{O}(\alpha_s^3)$ & $\mathcal{O}(\alpha_s^4)$ & $\mathcal{O}(\alpha_s^5)$ estimate \\ \hline  
    {\small $(\chi_1^2(BESIII)+\chi_1^2(KEDR))/ndf$}        &  
        11.445/25&
        10.958/25 &
        9.423/25& 
        10.419/25  \\
            P($\chi^2$), $\%$         &  
        99.050&
        99.315 &
        99.794 & 
        99.536  \\\hline \hline
        $\Lambda_{\rm{\overline{MS}}}^{(f=3)}$, MeV      &
        $ 374_{-113}^{+101}$  &  
        $ 420_{-136}^{+131}$   &  
        $ 618_{-162}^{+68}$   & 
        $ 518$\\ \hline
        $\nu$ (BES)         &  
        $0.966 \pm  0.019$  &  
        $0.969   \pm  0.020  $    
        & $ 0.964\pm  0.013 $ 
        & $0.955$     \\ \hline
        $\nu$ (KEDR)        & 
        $0.990 \pm  0.012$   &  
        $0.992 \pm 0.013$   & 
        $0.996 \pm 0.013$  & 
        $0.990$\\ \hline
        $\alpha_s(m_\tau)$  &   
        $0.3187_{-0.0556}^{+0.0564}$  &
        $0.3607_{-0.0725}^{+0.0836}$  &
        $0.5201_{-0.1305}^{+0.0769} $ & 
        $0.4203$\\ \hline
        $\alpha_s(M_Z)$ ($NLO)$&$0.1151_{-0.0069}^{+0.0052}$ &                      & & \\
        $\Lambda_{\rm{\overline{MS}}}^{(f=5)}$, MeV &$191_{-68}^{+65}$ &  &  &  \\
       \hline
        $\alpha_s(M_Z)$ ($NNLO)$&     &$0.1190_{-0.0081}^{+0.0064}$                  &  
       &\\
        $\Lambda_{\rm{\overline{MS}}}^{(f=5)}$, MeV &  & $220_{-84}^{+88}$  &  &  \\ \hline
        $\alpha_s(M_Z)$ ($N^3LO)$&                                 &    
        & $0.1283_{-0.0075}^{+0.0028}$ &  \\
        $\Lambda_{\rm{\overline{MS}}}^{(f=5)}$, MeV &  &  & $354_{-111}^{+49}$
        &  \\ \hline
        $\alpha_s(M_Z)$ ($N^4LO)$&  &                                 & & $0.1238$ \\
        $\Lambda_{\rm{\overline{MS}}}^{(f=5)}$, MeV &  &  &  & $284$ \\     \hline          
     \end{tabular}
     \end{table}
\end{center}
In our analysis of the BESIII data, we included only data points below the $J/\psi$ resonance, as 
incorporating  points above  it significantly  increases the $\chi^2$ value.
Restricting the BESIII dataset to this region enables a simultaneous fit to the data from both experiments. The resulting $\alpha_s$  values agree with the world average within experimental uncertainties and exhibit a steady increase with each perturbation order.
This upward trend in the extracted strong coupling constant stems from the structure of the higher-order contributions. Specifically, the perturbative series contain
$\pi^2$-terms generated by the analytic  continuation into Minkowski space. Since the corresponding coefficients grow in magnitude, truncating the series leads to progressively larger contributions.
In this study, we also evaluated quark-mass effects. A non-zero s-quark mass is expected to improve the convergence of the series at this energy scale, thereby flattening the upward trend in the extracted coupling constant.

\section{Summary and outlook}
We extracted the strong coupling constant using the low-energy data from
BESIII and KEDR.  There is a trend towards increasing values of the strong coupling constant as the perturbative order of the calculation increases. The experimental data are sufficiently precise to resolve effects associated with the specific nature of the theoretical calculations.
We also observe a noticeable influence of the strange quark mass on the calculated results.

\section*{Acknowledgements}
We thank Oleg Zenin for discussing and verifying several results.
\bibliographystyle{QCDEX2026}
\bibliography{proc2026} 

@article{Kataev:2026gea,
  author        = {A.~L.~Kataev and K.~Yu.~Todyshev},
  title         = {Perturbative QCD fitting of KEDR and BESIII $e^+e^-$ data for $R(s)$ and $\alpha_s(M_Z)$ determination},
  journal       = {Natural Sci. Rev.},
  volume        = {3},
  pages         = {100801},
  year          = {2026},
  doi           = {10.54546/NaturalSciRev.100801},
  eprint        = {2603.29803},
  archivePrefix = {arXiv}
}

@article{Anashin:2015woa,
  author        =  {V.~V.~Anashin and others},
  title         =  {Measurement of $R_{uds}$ and $R$ between 3.12 and 3.72 GeV at the KEDR detector},
  journal       = {Phys. Lett. B},
  volume        = {753},
  pages         = {533--541},
  year          = {2016},
  doi           = {10.1016/j.physletb.2015.12.059},
  eprint        = {1510.02667},
  archivePrefix = {arXiv}
}

@article{Anashin:2016hmv,
  author        = {V.V.Anashin  and others},
  title         = { Measurement of $R$ between 1.84 and 3.05 GeV at the KEDR detector},
  journal       = {Phys. Lett. B},
  volume        = {770},
  pages         = {174--181},
  year          = {2017},
  doi           = {10.1016/j.physletb.2017.04.073},
  eprint        = {1610.02827},
  archivePrefix = {arXiv}
}

@article{KEDR:2018hhr,
  author        = {V.~V.~Anashin and others},
  title         = {Precise measurement of $R_{uds}$ and $R$ between 1.84 and 3.72 GeV at the KEDR detector},
  journal       = {Phys. Lett. B},
  volume        = {788},
  pages         = {42--51},
  year          = {2019},
  doi           = {10.1016/j.physletb.2018.11.012},
  eprint        = {1805.06235},
  archivePrefix = {arXiv}
}

@article{BESIII:2021wib,
  author        = {M.~Ablikim and others},
  title         = {Measurement of the Cross Section for $e^{+}e^{-}\to$Hadrons at Energies from 2.2324 to 3.6710~GeV},      
  journal       = {Phys. Rev.  Lett.},
  volume        = {128},
  pages         = {062004},
  year          = {2022},
  doi           = {10.1103/PhysRevLett.128.062004},
  eprint        = {2112.11728},
  archivePrefix = {arXiv}
}

@article{Shen:2023,  
  author        = {J. M. Shen and others},
  title         = {Novel method to reliably determine the QCD coupling from R$_{uds}$ measurements and its effects to muon g{\ensuremath{-}} 2 and $\alpha \left({M}_Z^2\right)$ within the tau-charm energy region},
  journal       = {JHEP},
  volume        = {07},
  pages         = {109},           
  year          = {2023},
  doi           = {10.1007/JHEP07(2023)109},
  eprint        = {2303.11782},
  archivePrefix = {arXiv}
}

@article{Boito:2025qfz,
  author        = {D.~Boito and M.~Caram},
  title         = {Perturbative QCD below charm threshold: Theory and tensions with e+e- data},
  journal       = {Phys. Rev. D.},
  volume        = {112},
  no            = {9},
  pages         = {9},
  year          = {2025},
  doi           = {10.1103/dtg5-ghdx},
  eprint        = {2509.12956},
  archivePrefix = {arXiv}
}

@article{Marshall:1988ri,
  author        = {R. Marshall},
  title         = {A Determination of the Strong Coupling Constant $\alpha_s$ From $e^+ e^-$ Total Cross-section Data},
  journal       = {Z. Phys. C},
  volume        = {43},
  pages         = {595},
  year          = {1989},
  doi           = {10.1007/BF01550938}
}
\end{document}